\documentclass[a4paper,10pt,twoside]{cpc-hepnp}

\usepackage{multicol}
\usepackage{graphicx}
\usepackage{booktabs}
\usepackage{amssymb,bm,mathrsfs,bbm,amscd}
\usepackage[tbtags]{amsmath}
\usepackage{lastpage}
\usepackage{CJK}

\begin{document}

\fancyhead[co]{\footnotesize Huang Ming-Yang et al: The effects of injection beam parameters and foil scattering
for CSNS/RCS}


\title{The effects of injection beam parameters and foil scattering
for CSNS/RCS\thanks{supported
by National Natural
Science Foundation of China (Project Nos. 11175020 and 11175193)}}

\author{%
     Huang Ming-Yang$^{1)}$\email{huangmy@ihep.ac.cn}%
 \quad Wang Sheng%
 \quad Qiu Jing%
 \quad Wang Na%
 \quad Xu Shouyan
 } \maketitle

\address{%
(Institute of High Energy Physics, Chinese Academy of Sciences, Beijing 100049,
China)\\
}

\begin{abstract}
The China Spallation Neutron Source (CSNS) uses $H^-$ stripping
and phase space painting method to fill large ring acceptance with the linac
beam of small emittance.
The dependence of the painting beam on the injection beam parameters was studied
for the Rapid Cycling Synchrotron (RCS) of CSNS.
The injection processes for different momentum spread, $rms$ emittance of the injection beam,
injection beam matching were simulated, then the beam losses, 99\% and $rms$ emittances were obtained
and the optimized ranges of injection beam parameters were given.
The interaction between the $H^-$ beam and the stripping foil was studied and the foil scattering
was simulated. Then, the stripping efficiency
was calculated and the suitable thickness of the stripping foil was obtained.
The energy deposition on the foil and the beam losses due to the foil scattering were also studied.

\end{abstract}

\begin{keyword}
CSNS; RCS; Injection parameter; Foil scattering
\end{keyword}

\begin{pacs}
29.25.Dz, 29.27.Ac, 41.85.Ar, 34.50.Fa
\end{pacs}

\begin{multicols}{2}

\section{Introduction}

CSNS is a high power proton accelerator-based facility\cite{Wang1}.
The accelerator consists of a 1.6GeV RCS and an 80MeV $H^-$
linac which is upgradable to 250MeV. The RCS accumulates
$1.56\times10^{13}$ protons in two intense bunches
and operates at a 25Hz repetition rate with a design
beam power of 100kW, and is capable of upgrading to 500kW.
It has a four-fold lattice with four long straight sections
for the injection, extraction, RF and beam collimation.

For high intensity proton accelerators, injection via $H^-$
tripping is a practical method. The design
of the RCS injection system is to inject the $H^-$ beam into
the RCS with high precision and high transport efficiency.
In order to control the strong space charge effects which
are the main causes of the beam losses in CSNS/RCS, the phase
space painting method is used for injecting the beam of small emittance
from the linac into the large ring acceptance\cite{Tang1}.
With the code ORBIT\cite{Gabambos1}, the multi-turn phase space painting injection process
with space charge effects for CSNS/RCS is studied in detail.

When the $H^-$ beam traverses the
stripping foil, most of the particles $H^-$ are converted to $H^+$,
and the others are converted to $H^{\circ}$ or unchanged. The interaction with the stripping
foil can induce the beam emittance growth and beam loss. With the code FLUKA\cite{Ferrari1},
the foil scattering due to the interaction between the $H^-$ beam
and the stripping foil is simulated, and
the stripping efficiency is calculated.
The energy deposition on the foil and the beam losses due to the foil scattering are also studied.

\section{\label{sec:hamckm}Dependence of the painting beam on the injection beam parameters}

\begin{center}
\begin{tabular}{ccccc}
\scalebox{0.57}{\includegraphics{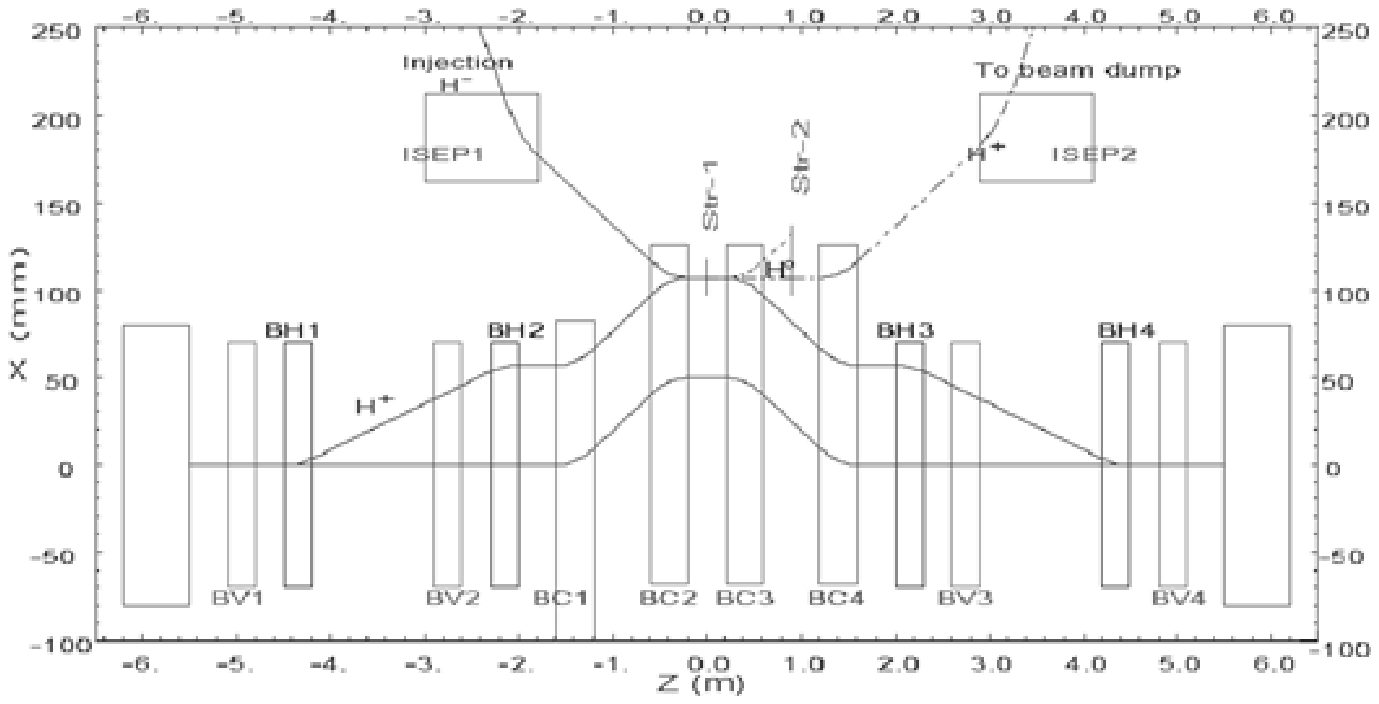}}\\
\end{tabular}\\
\figcaption{\label{injection}
Layout of the RCS injection system.}
\end{center}

For CSNS, combination of the $H^-$ stripping and phase space painting method
are used to accumulate high intensity beam in the RCS. Fig. 1 shows the layout of the RCS injection system\cite{Tang1} and Table 1 shows the main injection parameters\cite{CSNS1}.
For the beam injection, three kind of orbit-bump
systems are prepared\cite{Tang1}. One is a horizontal bump system (four dipole magnets; BH1-BH4)
for painting in $x-x'$ plane; one is a vertical bump system
(four dipole magnets; BV1-BV4) for painting in $y-y'$ plane; the third one is
a horizontal bump system (four dipole magnets; BC1-BC4) in the middle for an additional
closed-orbit shift of 47mm.

\begin{center}
\tabcaption{ \label{tab1}  Main injection parameters of CSNS/RCS.}
\footnotesize
\begin{tabular*}{80mm}{c@{\extracolsep{\fill}}c}
\toprule Parameters/units & Values   \\
\hline
Circumference/m & 227.92  \\
Injection energy/GeV & 0.08  \\
Extraction energy/GeV & 1.6 \\
Injection beam power/kW  &  5  \\
Extraction beam power/kW  &  100  \\
Nominal betatron tunes   & 4.86/4.78 \\
RF frequency/MHz  &  $1.0241\sim2.3723$ \\
RF voltage/kV & 165 \\
Harmonic number  &  2  \\
Repetition rate/Hz  &  25 \\
Number of particles per pulse  &  $1.56\times10^{13}$  \\
Momentum acceptance  &  1\%  \\
Painting scheme  &  Anti-Correlated  \\
Chopping rate   &  50\%  \\
Turn number of injection  &  200  \\
\bottomrule
\end{tabular*}
\end{center}

\begin{center}
\begin{tabular}{ccccc}
\scalebox{0.5}{\includegraphics{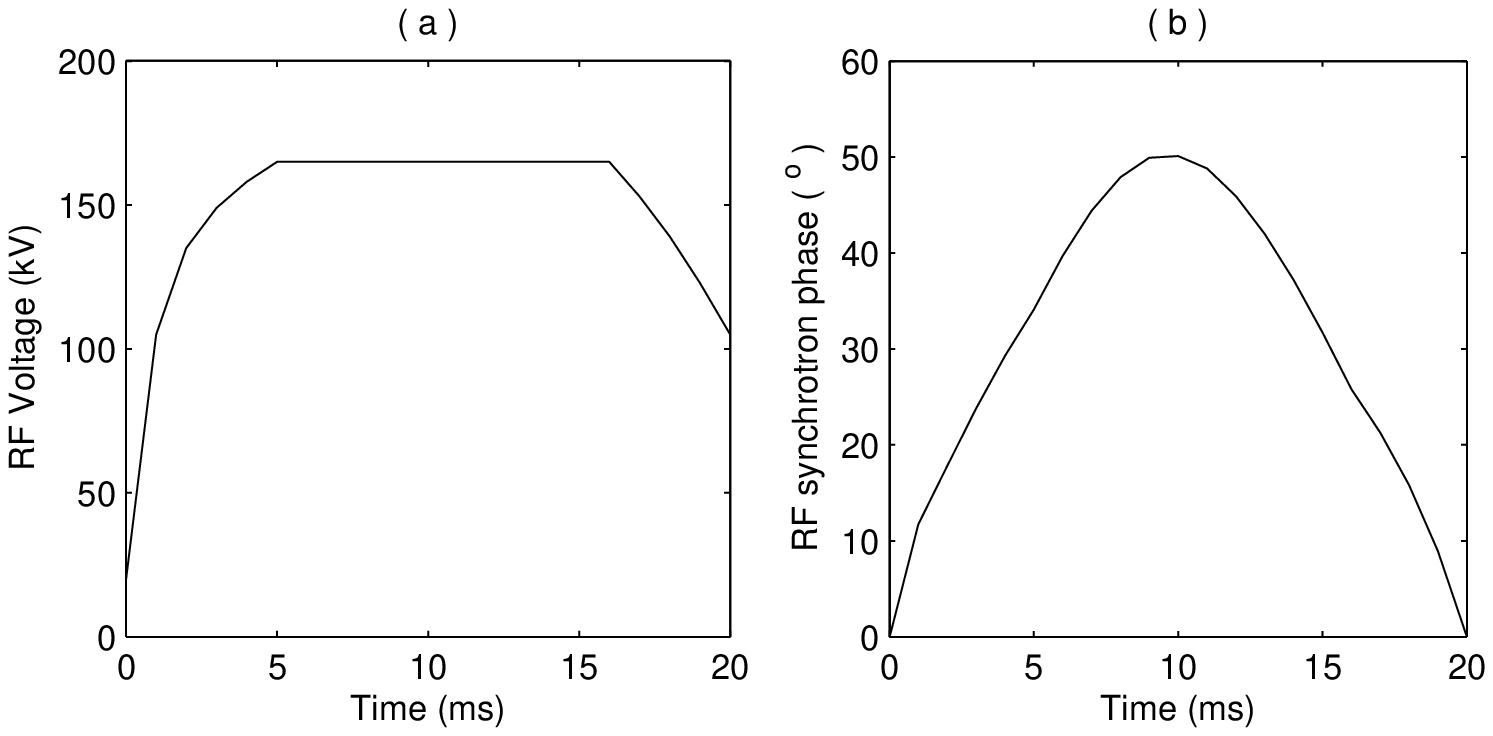}}\\
\end{tabular}\\
\figcaption{\label{RF}
The patterns of the RF voltage and synchronous phase over the acceleration period.}
\end{center}

In the RCS, the emittance evolution and beam losses depend on the injection beam
parameters, such as the injection emittance, starting injection time,
injection beam matching, momentum spread,
and chopping rate.
Some works have been done for the injection
parameters optimization\cite{Qiu1}\cite{Weit1}. In this section,
the effects of the momentum
spread, the $rms$ emittance of the injection beam,
and the injection twiss parameters mismatch are discussed in detail.
In the falling simulation, the
chopping rate is $50\%$,
the patterns of the RF voltage and synchrotron phase are given in Fig. 2,
and the space charge effects are considered. At the same time,
the turn number of the injection painting process is 200 and only 2000
turns in the acceleration process are considered for the simulation.

\subsection{\label{sec:hamckm}Momentum spread}

\begin{center}
\begin{tabular}{ccccc}
\scalebox{0.6}{\includegraphics{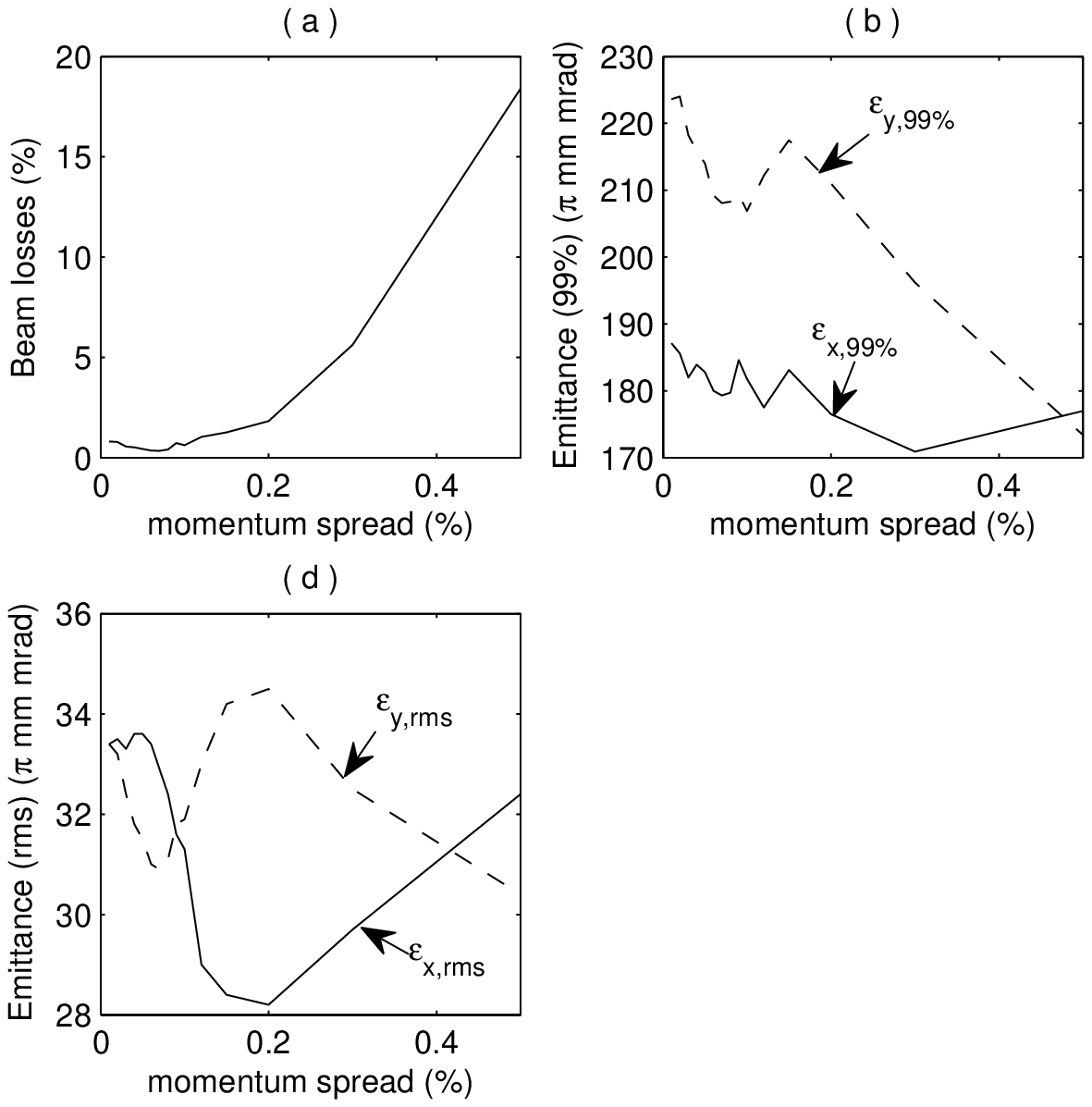}}\\
\end{tabular}\\
\figcaption{\label{mom}
Beam losses, 99\% and $rms$ emittances as a function of the momentum spread.}
\end{center}

The code ORBIT is used for the injection simulation, it can perform the painting
injection process and include the space charge effects.
By using ORBIT, the injection processes
with the momentum spread between $0.01\%$ and $0.5\%$ were simulated.
Fig. 3 shows the beam losses, 99\% and $rms$ emittances as a function of the momentum spread
and Fig. 4 shows the $rms$ emittance evolution for different momentum spread.
It can be found from Fig. 3 that the beam losses decrease firstly and then
increase with the increasing momentum spread. While the momentum
spread smaller than $0.1\%$, the beam losses are smaller than
$1\%$, the 99\% and $rms$ emittances are constrained in reasonable ranges.

It can be found
from Fig. 4 that there is transverse coupling between $x$ and $y$
$rms$ emittance evolutions which depends on the momentum
spread. When the momentum spread is below $0.1\%$, the coupling
becomes stronger and stronger with the increasing momentum spread.
However, when the momentum spread is above $0.1\%$, the coupling
becomes weaker and weaker with the increasing momentum spread.
Therefore, the momentum spread of $0.1\%$ is a optimal value
for the injection.
This simulation results are consistent with the running experiments in J-PARC\cite{Wei1}.

\begin{center}
\begin{tabular}{ccccc}
\scalebox{0.6}{\includegraphics{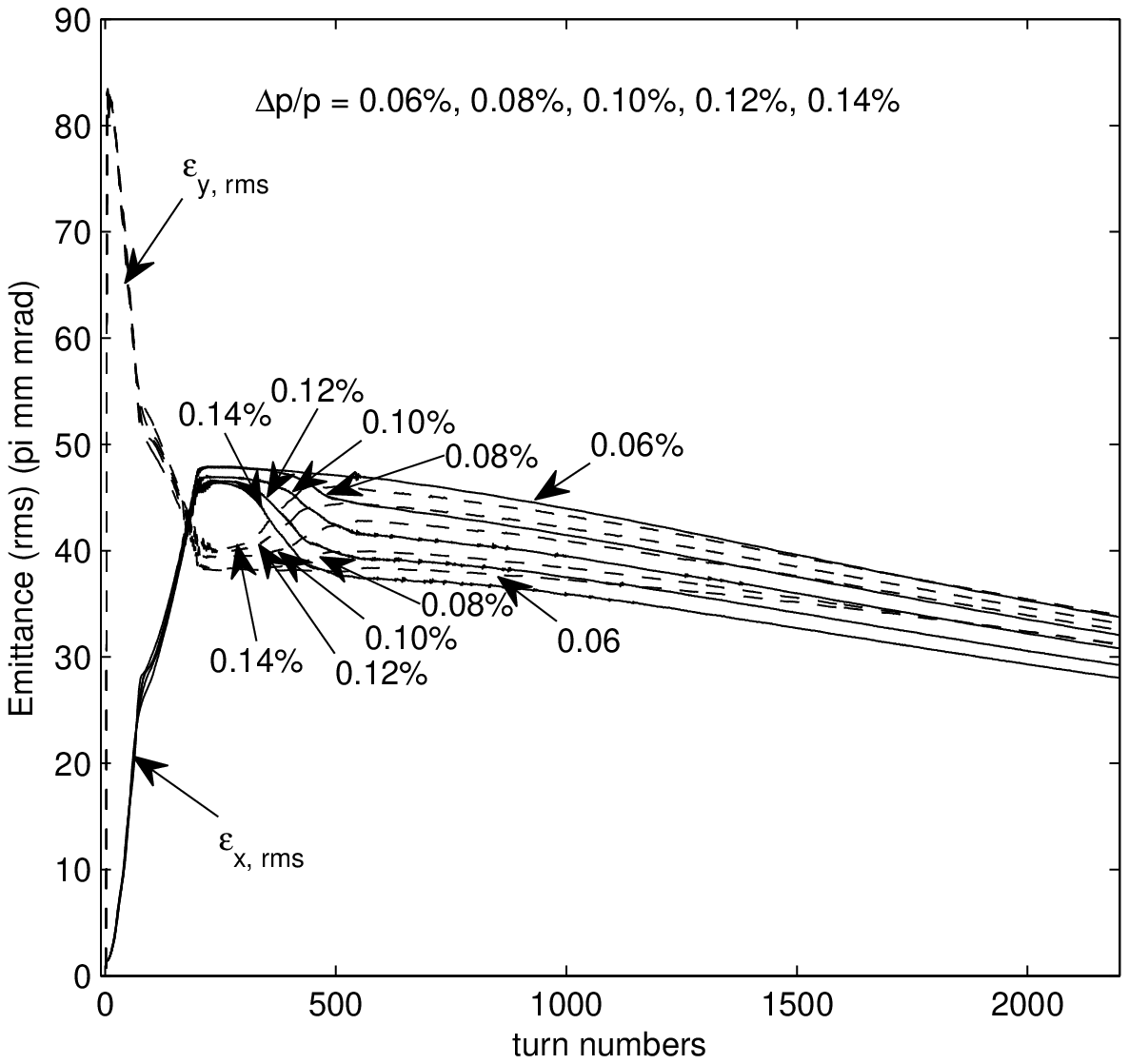}}\\
\end{tabular}\\
\figcaption{\label{coupling1}
The $rms$ emittance evolution for different momentum spreads.}
\end{center}

\subsection{\label{sec:hamckm}$rms$ emittance of the injection beam}

\begin{center}
\begin{tabular}{ccccc}
\scalebox{0.42}{\includegraphics{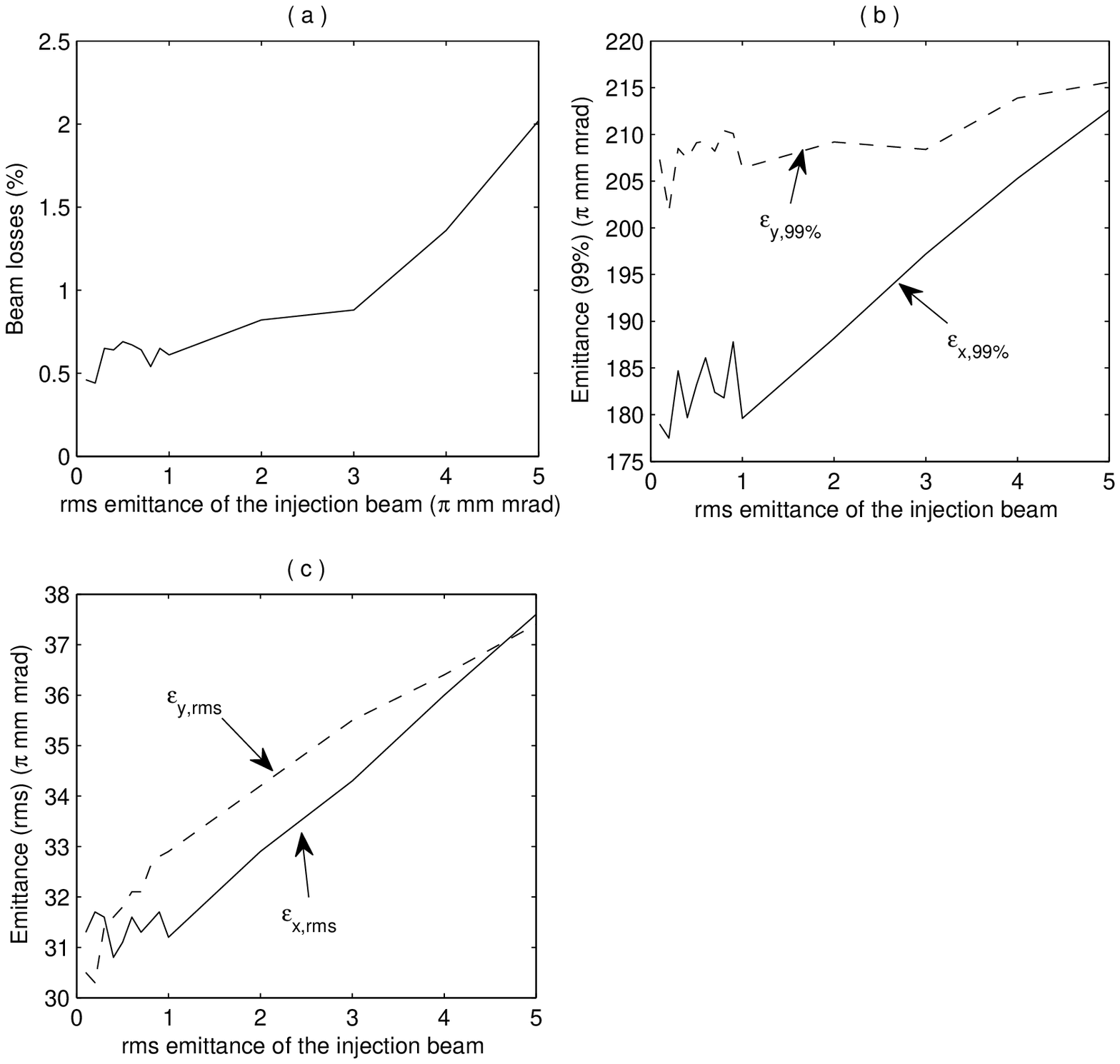}}\\
\end{tabular}\\
\figcaption{\label{mom}
Beam losses, 99\% and $rms$ emittances as a function of the $rms$ emittance of the injection beam.}
\end{center}

In order to study the effects of the $rms$ emittance of the injection beam,
the injection processes with the
$rms$ emittance between $0.1\pi mm\cdot mrad$ and $5.0\pi mm\cdot mrad$ were simulated.
Fig. 5 shows the beam losses, 99\% and $rms$ emittances as a function of the
$rms$ emittance of the injection beam. It can be found that the beam losses, $99\%$ and $rms$ emittances
all increase with the increasing $rms$ emittance of the injection beam. In addition, while
the $rms$ emittance of the injection beam is smaller than $1.0\pi mm\cdot mrad$,
the beam losses are smaller than $1\%$, the $99\%$ and $rms$ emittances
are constrained in reasonable ranges.

\subsection{\label{sec:hamckm}Injection twiss parameters mismatch}

For the RCS design, it has been a primary concern to match the physical parameters of the linac
and the RCS at the injection point. A mismatched injection could result in large beam losses
and an undesirable transverse emittance growth. The first condition for the injection beam matching
is obtained by choosing the parameters\cite{Beebe1}:
\begin{eqnarray}
 \frac{\alpha_l}{\beta_l}=\frac{\alpha_r}{\beta_r},
\label{match}
\end{eqnarray}
where $\alpha_l$ and $\beta_l$ are the twiss parameters for the linac and
$\alpha_r$ and $\beta_r$ for the RCS. For CSNS, $\alpha_r$ nearly equals to 0.
In order to study the effects of the injection twiss
parameters mismatch, for a fixed $\beta_l$, the injection processes for different
$\alpha_l$ were discussed.

\begin{center}
\begin{tabular}{ccccc}
\scalebox{0.52}{\includegraphics{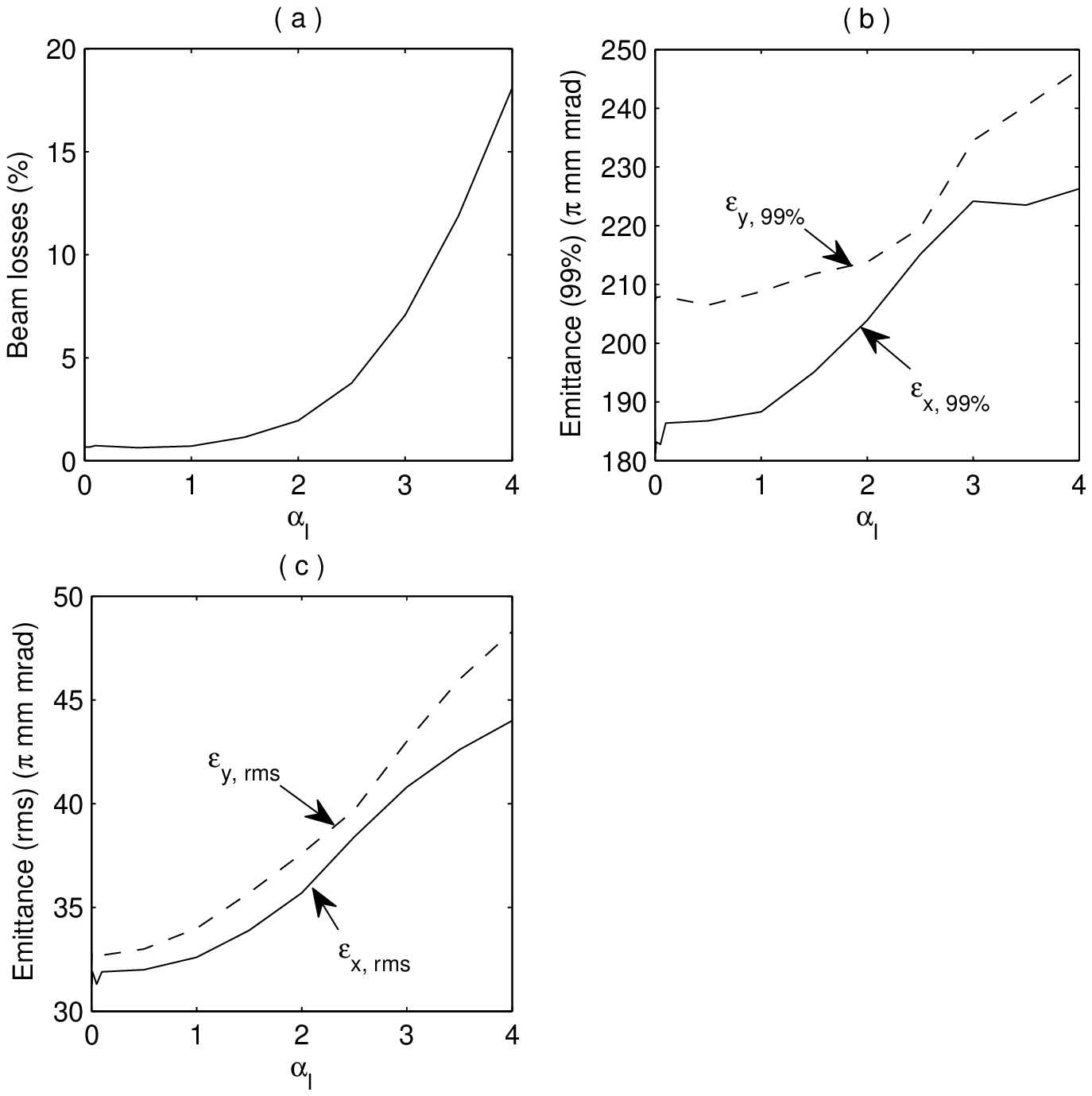}}\\
\end{tabular}\\
\figcaption{\label{mom}
Beam losses, 99\% and $rms$ emittances as a function of $\alpha_l$.}
\end{center}

The injection processes with
$(\alpha_{lx}, \alpha_{ly})$ between (0.0, 0.0) and (5.0, 5.0) were simulated.
Fig. 6 shows the beam losses, 99\% and $rms$ emittances as a function of $\alpha_l$.
It can be found that the beam losses, $99\%$ and $rms$ emittances
all increase with the increasing $\alpha_l$. While $(\alpha_{lx}, \alpha_{ly})$ smaller
than (1.0, 1.0), the beam losses are
smaller than $1\%$, the 99\% and $rms$ emittances
are constrained in reasonable ranges, i.e. the effects of the
injection twiss parameters mismatch are very small.
However, while $(\alpha_{lx}, \alpha_{ly})$ larger than
(1.0, 1.0), the beam losses, $99\%$ and $rms$ emittances are much larger than
that of the match case, i.e. the injection beam should not be matched into the RCS
acceptance.

In the above discussions, we have studied the effects of the momentum spread,
the $rms$ emittance of the injection beam, and the injection twiss parameters mismatch.
It can be found that
the beam losses are smaller than $1\%$, the 99\% and $rms$ emittances
are constrained in reasonable ranges while the momentum spread smaller
than $0.1\%$, the $rms$ emittances of the injection beam
smaller than $1.0\pi mm\cdot mrad$, and $(\alpha_{lx}, \alpha_{ly})$ smaller than (1.0, 1.0).
The momentum spread of $0.1\%$ is a optimal value
for the injection, and the injection beam should be well matched
into the RCS acceptance when $(\alpha_{lx}, \alpha_{ly})$ smaller than (1.0, 1.0).

\section{\label{sec:hamckm}Foil scattering effects}

In the injection system of the RCS, there are two carbon stripping foils,
a primary stripping foil and a secondary stripping foil.
In this section, the interaction between the $H^-$ beam and
the primary stripping foil is discussed\cite{Mohagheghi1}\cite{Saha2}\cite{Drozhdin1} and
the stripping efficiency is calculated. The energy deposition on the foil and the beam losses
due to the foil scattering are also studied.

\subsection{\label{sec:hamckm}Foil stripping}

\begin{center}
\begin{tabular}{ccccc}
\scalebox{0.7}{\includegraphics{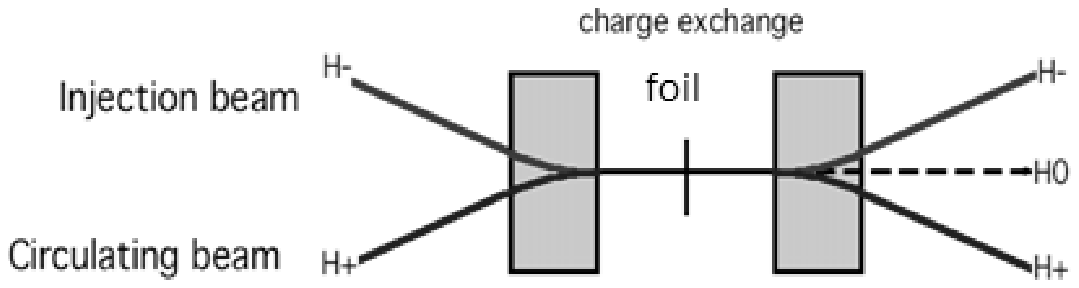}}\\
\end{tabular}\\
\figcaption{\label{foil}
The production of $H^-$, $H^0$, $H^+$ by foil stripping.}
\end{center}

When the $H^-$ beam traverses the carbon stripping foil\cite{Gulley1},
there are six  charge exchange processes: three are electron loss reactions
and three are electron pickup reactions. However, for energies above 100keV,
the cross sections for electron pickup are very small and can be neglected.
Therefore, the remain particles after foil stripping are $H^-$, $H^{\circ}$
and $H^+$, as shown in Fig. 7. The
stripping efficiency of $H^+$ is given by\cite{Webber1}
\begin{eqnarray}
 f_{H^+}=1-\frac{1}{\sigma_{-1,0}+\sigma_{-1,1}-\sigma_{0,1}}\Bigg[\sigma_{-1,0}e^{-\sigma_{0,1}x} \nonumber\\ -(\sigma_{0,1}-\sigma_{-1,1})e^{-(\sigma_{-1,0}+\sigma_{-1,1})x}\Bigg], \label{fh1}
\end{eqnarray}
where $\sigma_{-1,0}$, $\sigma_{0,1}$, $\sigma_{-1,1}$ are the cross-sections of the
reactions $H^-\rightarrow H^{\circ}+e^-$, $H^{\circ}\rightarrow H^++e^-$, and
$H^-\rightarrow H^++e^-+e^-$, respectively. In addition, $x=N_{\circ}\tau/A$, where
$N_{\circ}$ is the Avogadro's constant, $A$ is the atomic number of the carbon foil,
and $\tau$ is the area density. The percent of the $H^-$ beam traverses the carbon foil
without stripping is given by\cite{Webber1}
\begin{eqnarray}
 f_{H^-}=e^{-\sigma_{-1,0}x}.
\label{fh2}
\end{eqnarray}
Therefore, the yielding rate of $H^{\circ}$ can be expressed as
\begin{eqnarray}
 f_{H^\circ}=1-f_{H^+}-f_{H^-}.
\label{fh3}
\end{eqnarray}
There are some studies\cite{Webber1}\cite{Chou1} about the cross-sections $\sigma_{-1,0}$, $\sigma_{0,1}$,
$\sigma_{-1,1}$ which depend on the beam energy.
Table 2 shows a summary of the cross-sections at 80MeV and 250MeV.

\begin{center}
\tabcaption{ \label{tab2}  Cross-sections of $H^-$ incident on carbon foil (unit $10^{-18}cm^2$).}
\footnotesize
\begin{tabular*}{80mm}{c@{\extracolsep{\fill}}cc}
\toprule  & 80MeV    & 250MeV   \\
\hline
$\sigma_{-1,0}$ & 3.17  & 1.35  \\
$\sigma_{0,1}$  & 1.24  & 0.53  \\
$\sigma_{-1,1}$ & 0.056 & 0.024 \\
\bottomrule
\end{tabular*}
\end{center}

\begin{center}
\begin{tabular}{ccccc}
\scalebox{0.48}{\includegraphics{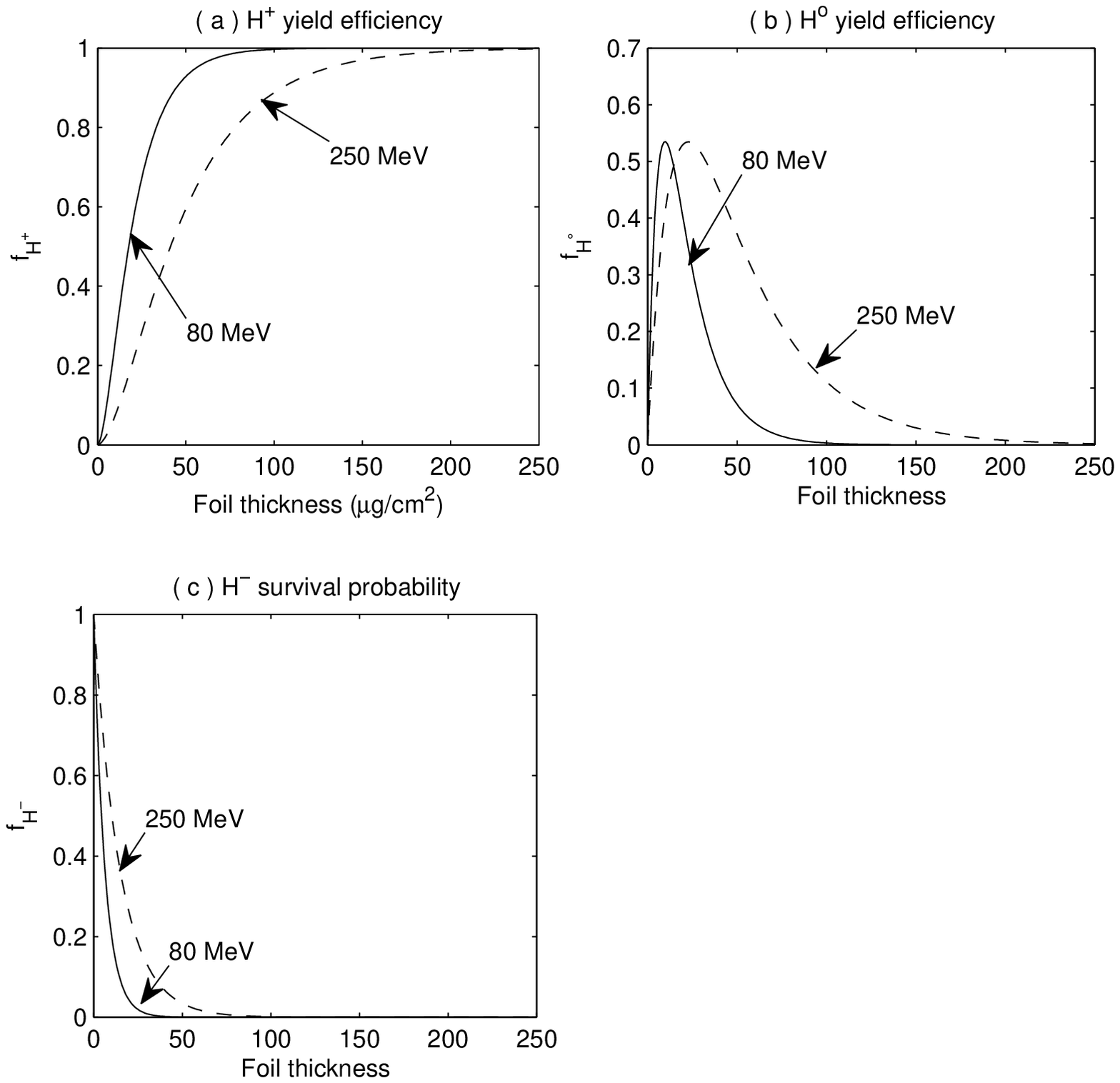}}\\
\end{tabular}\\
\figcaption{\label{pain}
$H^-$, $H^0$, $H^+$ yielding rates as a function of the foil thickness.}
\end{center}

Using Eqs. (\ref{fh1})- (\ref{fh3}) and the cross-sections given in Table 2,
the relations between $f_{H^+}$, $f_{H^{\circ}}$, $f_{H^-}$ and the foil thickness
can be obtained. Fig. 8 shows the curves that $f_{H^+}$, $f_{H^{\circ}}$, $f_{H^-}$
vary with the foil thickness. It can be found that, with the increasing
thickness of the foil, $f_{H^+}$ increases, $f_{H^-}$ decreases, and $f_{H^{\circ}}$
has a maximum value. For a given foil thickness, the stripping efficiency for 80MeV injection is larger than
that for 250MeV injection. For CSNS/RCS injection system, in order to make the stripping
efficiency greater than $99.7\%$, the thickness of the stripping foil need to be larger than
$100\mu g/cm^2$ for  80MeV injection and $240\mu g/cm^2$ for 250MeV injection.

\subsection{\label{sec:hamckm}Energy deposition on the foil}

\begin{center}
\begin{tabular}{ccccc}
\scalebox{0.4}{\includegraphics{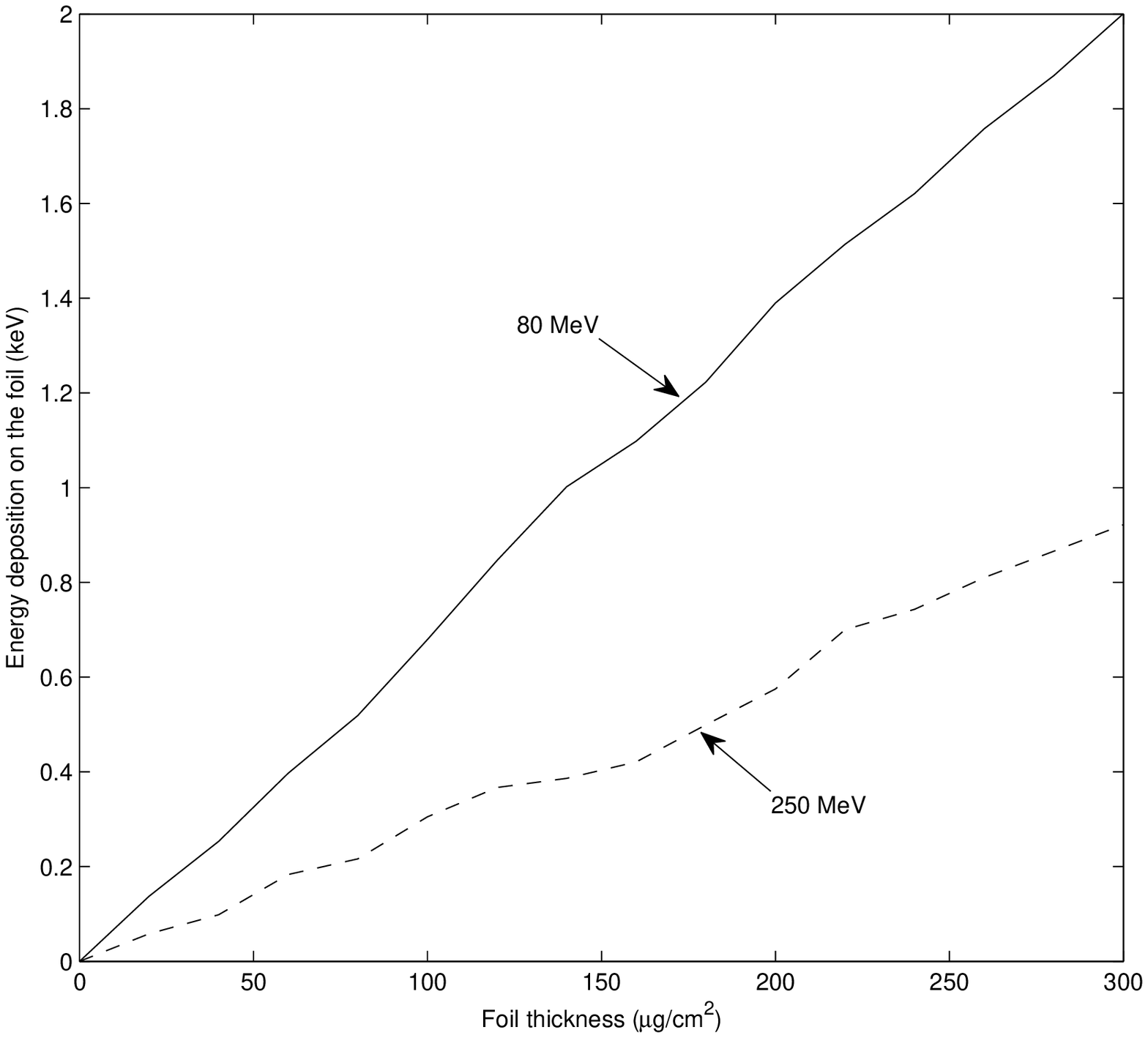}}\\
\end{tabular}\\
\figcaption{\label{energy}
The energy deposition as a
function of the foil thickness.}
\end{center}

When the $H^+$ beam traverses the stripping foil,
there is an energy deposition on the foil due to the foil scattering.
The energy deposition depends on the beam energy
and the foil thickness\cite{Drozhdin1}. In this part, the relations between the
energy deposition and the thickness of the stripping foil for different injection
energy are studied.

By using the code FLUKA, the foil scattering process were simulated when the
$H^+$ beam traverses the stripping foil.
Fig. 9 shows the energy deposition on the foil as a function of the thickness of the stripping foil.
Both for 80MeV injection and 250MeV injection, it can be found that the energy
deposition increases with the increasing thickness of the stripping foil.
In addition, the relations between the energy deposition and the foil thickness
are nearly linear. Furthermore, the energy deposition on the foil is 0.68keV for
80MeV injection ($100\mu g/cm^2$) and 0.74keV for 250MeV injection ($240\mu g/cm^2$).

\subsection{\label{sec:hamckm}Beam losses}

During the injection process, the foil scattering will generate the beam
halo and result in additional beam losses.
For J-PARC, the stripping foil scattering has been studied and
it can be found that the foil scattering
is the main cause of beam losses in the injection region\cite{Akasaka1}.
Therefore, the beam losses due to the foil scattering for CSNS/RCS also
need to be studied in detail. By using the codes
FLUKA and ORBIT, the injection process
and foil scattering can be simulated. Table 3 shows the beam parameters
for 80MeV
injection and 250MeV injection.

\begin{center}
\tabcaption{ \label{tab3}  Beam parameters for 80MeV
injection and 250MeV injection.}
\footnotesize
\begin{tabular*}{80mm}{c@{\extracolsep{\fill}}cc}
\toprule Injection & 80MeV    & 250MeV  \\
\hline
Injection beam power/KW           &  5     & 80     \\
Average injection current/$\mu A$ &  62.5  & 312.5  \\
Turn number of injection          &  200  &  403  \\
Foil thickness/($\mu g/cm^2$)       &  100   & 240    \\
\bottomrule
\end{tabular*}
\end{center}

By using the code ORBIT, the injection process can be simulated,
the average traversal number and the beam distribution after injection can be obtained.
Calculating those particles of the beam distribution which are in the range of the stripping foil,
the twiss parameters and $99\%$ emittance for those
particles can be obtained, as shown in Table 4. With these beam parameters,
the beam distribution that hitting on the stripping foil can be simulated.
Then, the foil scattering process can be simulated by the code FLUKA and the beam
losses due to the foil scattering in single turn
can be obtained. By using the average traversal number, the foil scattering induced beam losses
during the multi-turn injection process can be calculated.
Table 5 shows a summary of the beam losses due
to the foil scattering.
It can be found
that the beam losses are about 0.3W for 80MeV
injection and 4.6W for 250MeV injection.

\begin{center}
\tabcaption{ \label{tab3}  Beam parameters of
the proton distribution that hitting on the stripping foil.}
\footnotesize
\begin{tabular*}{80mm}{c@{\extracolsep{\fill}}cc}
\toprule Injection & 80MeV    & 250MeV  \\
\hline
$(\alpha_{x}, \alpha_{y})$            & (0.003, 0.044)   & (0.001, 0.016)  \\
$(\beta_{x}, \beta_{y})$/m            & (1.833, 4.458)   & (1.877, 5.222)  \\
$(\gamma_{x}, \gamma_{y})$/$m^{-1}$   & (0.546, 0.225)   & (0.533, 0.192) \\
$(\varepsilon_{x,99\%}, \varepsilon_{y,99\%})$/$(\pi \cdot mm \cdot mrad)$ & (92, 247)   & (90, 282) \\
\bottomrule
\end{tabular*}
\end{center}

\begin{center}
\tabcaption{ \label{tab3}  Beam losses due to the stripping foil scattering.}
\footnotesize
\begin{tabular*}{80mm}{c@{\extracolsep{\fill}}cc}
\toprule Injection & 80MeV    & 250MeV  \\
\hline
Average traversal number   &  5   &   10  \\
Particle loss ratio in single turn     & 0.0012\%  & 0.00058\%  \\
Total beam losses/W                 & 0.3   & 4.6 \\
\bottomrule
\end{tabular*}
\end{center}

\section{\label{sec:hamckm}Conclusions}

The dependence of the painting beam on the injection beam parameters for CSNS/RCS were studied and
the injection processes for different momentum spread,
$rms$ emittance of the injection beam, and injection beam matching had been discussed.
The beam losses, 99\% and $rms$ emittances were obtained, and the optimized ranges of
injection beam parameters were given.
The interaction between the $H^-$ beam and the stripping foil was studied.
Then, the stripping efficiency of $H^+$ and the yielding rates
of $H^-$ and $H^{\circ}$ were calculated. The energy deposition on the foil and
the beam losses due to the foil scattering were also studied.

\vspace{5mm}

\acknowledgments{The authors would like to thank CSNS colleagues for the discussions
and consultations.}

\end{multicols}

\vspace{-2mm}
\centerline{\rule{80mm}{0.1pt}}
\vspace{2mm}

\begin{multicols}{2}

\end{multicols}

\end{document}